# The local balance laws for energy, momentum and entropy: how they came into being, and what was their destiny

Friedrich Herrmann

Physics Education Group, Institute of Theoretical Solid State Physics, Karlsruhe Institute of Technology, D-76128 Karlsruhe, Germany

f.herrmann@kit.edu

**Abstract:** The historical process of the genesis of the extensive or substance-like quantities took place in two steps. First, global conservation or non-conservation was discovered. Only later did it become possible to formulate the balance locally in the form of a continuity equation. This process can be clearly seen in energy, momentum, and entropy. After a long and intricate history, the quantitative description of the local balance has been achieved for all of the three quantities in a surprisingly short period of time around the turn of the 19th to the 20th century. The new ideas could have simplified considerably the teaching of energy, momentum, and entropy. However, in all three cases, today's language of physics remained essentially the same as it was at the time when a local balancing was not yet possible.

## 1 Introduction

There are physical quantities which have the property to measure the amount of "something": Momentum measures an amount of motion, mass is a measure of an "amount of inertia", entropy measures the "amount of heat" (in a colloquial sense) (Job, p.45 and Rüffler 2011, Fuchs 2010, p 3 and pp 116–7); some others, such as energy and electric charge, are measures of the amount of something for which we do not have an immediate perception. Since the value of each of these quantities refers to a region of space, they are called extensive quantities. We can say they measure a bulk property.

Compared to other physical quantities their handling is particularly simple. When dealing with an extensive quantity, one can make use of a powerful model: the *substance model*. One imagines the quantity as a measure of the amount of a substance or fluid, and one speaks about the quantity as if one were speaking about this imaginary substance. Because of this possibility the extensive quantities are also called substance-like quantities (Falk et al. 1983, Herrmann 2000) or fluid-like quantities (Fuchs 2010, p 9).

When dealing with a substance-like quantity, we do something that can in colloquial terms be called "balancing" or "accounting". The change of the value of the quantity $X$ in a given region of space is linked to an inflow or outflow, or to the production or destruction of $X$.

Balancing or accounting is a simple task. The math that is needed is just adding and subtracting. Every child is able to do these operations and does so when dealing with objects of our daily life: with people, with water, with sand or jelly babies, but also with abstract concepts like money.

When teaching physics, at school and at university, this simple property of the extensive quantities does not always become clear. One does not always benefit from it. Only mass and electric charge are consistently treated in such a way that the substance or fluid character is expressed and exploited.

Although in daily life energy is treated as a commodity, i.e. the substance model is used, in physics it is still common to use a language that hinders an understanding of the extensive character of the quantity; for example, when saying that "work is done upon a body" or "work is performed" during a mechanical energy transport, or when talking about the "power" of a device, instead of the energy current that flows into (and out of) the device.

As far as momentum is concerned, in the context of collision experiments, it is treated like an amount of something that is transmitted. However, the transmission process itself is hidden in the so-called law of interaction, which is formulated in such a way that the question of the path along which the momentum passes from one body to another does not arise.

For entropy things are even worse. If it is introduced as a measure of the value of energy, as a measure of the irreversibility of a process, or as a measure of microscopic disorder the idea of a quantity for which a balance can be established does hardly show up.

Why don't we take advantage of the simple properties of these three physical quantities – energy, momentum and entropy? The answer to this question can be found in their genesis. All the three had a turbulent history marked by several adversities.

However, one thing is striking. Although the history of their origins goes back several hundred years and is very different for the three quantities, the proof that a local balance can be established was provided within a period of time of less than 30 years, namely between 1884 and 1911, i.e. within a single generation of researchers. There was a real breakthrough. Apparently the time had arrived.

I will remind in section 2 what is meant by a balance equation. In section 3 the history of the three quantities and the formulation of the non-local, provisional balance laws is briefly traced. In section 4 I sketch the short period of time when the local balance laws finally were formulated. In section 5, I examine the effects of the new formulation on our actual way of teaching physics and in section 6 we will see what remains to be done.

## 2 Balance equations

For each substance-like quantity $X$ an equation of the form

$$\frac{dX}{dt} = I_X + \Sigma_X \tag{1}$$

can be formulated.

The equation can be interpreted as follows: The time rate of change $dX/dt$ of $X$, comes about in two ways:

- by an inflow or outflow, expressed by the current intensity $I_X$

- by production or destruction, expressed by the production rate $\Sigma_X$ (produced amount of $X$ by time).

If $X$ is "destroyed", $\Sigma_X$ is negative.

We call equation (1) the balance equation of $X$.

If for a quantity $X$ the term $\Sigma_X$ is always zero, $X$ is a conserved quantity. The balance equation then reads:

$$\frac{dX}{dt} = I_X \tag{2}$$

In this case, the value of $X$ can only change by means of an inflow or outflow.

Equations (1) and (2) make a statement about a region of space. $dX/dt$ and $\Sigma_X$ refer to the inside of the region, $I_X$ to its surface. The equations are "integral" formulations of the balance of $X$. It is also possible to express a balance "locally", i.e. with quantities that refer to a point:

$$\frac{\partial \rho_X}{\partial t} + \operatorname{div} \mathbf{j}_X = \sigma_X \tag{3}$$

Here $\rho_X$ is the density of $X$, $\mathbf{j}_X$ is the current density and $\sigma_X$ is the production rate per volume. Equation (1) follows from equation (3) by integration over the volume of the considered region of space. A relation of the form of equation (3) is called a continuity equation or local balance equation.

In the case that $X$ is the electric charge $Q$, the local balance equation reads:

$$\frac{\partial \rho_Q}{\partial t} + \operatorname{div} \mathbf{j}_Q = \sigma_Q$$

Here $\rho_Q$ is the electric charge density and $\mathbf{j}_Q$ the electric current density. The production term is zero, because electric charge is a conserved quantity.

## 3 Weak balance laws

A physical quantity usually "originates" little by little. We tend to ask for the inventor or discoverer of a physical quantity, and sometimes we believe we can identify one: The energy was introduced by Mayer and Joule, the entropy by Clausius, momentum comes from Huygens, electric charge from Franklin. Actually, these researchers have made important contributions to the "construction" of the respective quantity, but in all of these cases the actual history of origin begins earlier, and it was not finished with the contributions of the aforementioned scientists.

In several cases the quantity was first introduced long before it became possible to recognize that a local balance could be established.

Typically, a substance-like quantity $X$ was introduced after discovering that the sum of its values on two systems A and B is either constant or, as in the case of entropy, constant or increasing in time:

$X_A + X_B$ = const,

or

$X_A + X_B$ = const or increasing

respectively.

Notice that there is no flow or current in such a formulation. No statement is made about what happens outside the systems A and B. Nothing is said about the role which a medium between A and B might play. Nothing is said about a transport of $X$ between A and B. In principle, this kind of formulation suggests the existence of an action at a distance.

I will call such a proposition, which only expresses the constancy of the total amount, a "weak balance law". Correspondingly, a balance law, in which the transport process is included, is a "strong balance law". If a strong balance law applies, the process can be described with a continuity equation.

Let us now have a look at the early history of our three extensive quantities. We begin with momentum.

## 3.1 Momentum

The quantity has a long history, and it can be said with good reason that Galileo, Déscartes, and Huygens were already familiar with simple versions of the principle of momentum conservation. Here, we will limit ourselves to Newton's version, for it was Newton who was concerned with the exchange of momentum by means of gravity, and this represented the greatest problem in the mathematical formulation of the momentum balance.

Newton described what we now call the conservation of momentum by means of his three laws. Let us formulate Newton's second law in modern language:

"The time rate of change of momentum of a body is proportional to the force that acts on the body."

Without doing violence to the statement, we can pour it into a formula, with the symbols commonly used today:

$$\frac{d\mathbf{p}}{dt} = \mathbf{F} \tag{4}$$

Since the value of the momentum refers to a region of space, and since the equation has the form of equation (2), it can be read as a balance law. Does that mean that Newton already had the momentum balance law in a modern sense? No, it doesn't. He was not yet able to interpret the quantity **F** as a flow of momentum. If **F** is a flow, one has to be able to specify the path that the momentum takes when going from one body to another: from a magnetic pole to a piece of soft iron, or from the earth to the moon. But that was not possible in Newton's time. Today we would say: the field concept did not yet exist.

We can also ask: If **F** is a flow, its value refers to a sectional surface between the two bodies. But what does Newton's force refer to? We all learned it in school and at university: to the two bodies themselves – to the body which exerts the force, and to that on which the force is exerted.

This way of speaking about equation (4) is an ingenious invention of Newton. The Newtonian wording only refers to the two bodies; the medium between them is not mentioned. The question of what role such a medium might play does not arise. Newton makes no allusion to what might exist in between, that is, to what today we call a field.

One might ask: But was Newton so naive to believe in actions at a distance? He commented the subject in an appendix ("General Scholium") to the second edition of his Principia:

> "But hitherto I have not been able to discover the cause of those properties of gravity from phenomena, and I frame no hypotheses; for whatever is not deduced from the phenomena is to be called an hypothesis, and hypotheses, whether metaphysical or physical, whether occult qualities or mechanical, have no place in experimental philosophy."

Newton did not want his stringent work to contain speculative elements.

It is a different question what his personal conviction was regarding the actions at a distance. One learns it from a letter to the English scholar Richard Bentley:

> "That gravity should be innate, inherent and essential to Matter, so that one Body may act upon another at a Distance thro' a Vacuum, without the Mediation of any thing else, by and through which their Action and Force may be conveyed from one to another, is to me so great an Absurdity, that I believe no Man who has in philosophical Matters a competent Faculty of thinking, can ever fall into it. Gravity must be caused by an Agent acting constantly according to certain Laws; …."

One hardly could say it more clearly. Apparently, Newton was convinced that one day a strong balance law could be formulated.

Nevertheless, Newton's "weak" version of the balance equation remained state of the art for more than 150 years. Thereafter, other substance-like quantities entered the stage, and also for these only weak balance laws could initially be formulated.

### 3.2 Energy

Forerunners of the energy, such as the "vis viva" (living force), the "motive power" and "work", existed before the energy was introduced as a quantity of its own right. The years from 1841 through 1847 are usually considered the time interval in which the physical quantity energy was born – through the work of Robert Mayer, James Prescott Joule and Hermann von Helmholtz.

Again, at first one could only confirm that the increase in the value of the newly introduced quantity in one system is linked to the decrease in another one.

As Mayer (1874) puts it:

> "In all physical and chemical processes, the given force remains a constant quantity."

By "force" he means what we call energy today.

Once again, only a weak balance law could be formulated.

But also in this case there was an early expectation that the new quantity should be conserved locally. The young Max Planck (1908) nicely describes this as an yet unconfirmed conjecture in 1887 in the booklet "The Principle of the Conservation of Energy":

> "One will no longer be satisfied with knowing the numerical value of the energy of the system, but one will try to show the existence of the various kinds of energy at the different elements of the system in detail, and to follow the transition... to other elements just as the movement of a quantum of matter in space.... Currently it is in any case a matter of physical research to elaborate this view in detail as the most descriptive and fruitful one and to examine its consequences on the basis of experience …"

In 1892 Heinrich Hertz also expressed the uncertainty that still existed in this regard:

> "A greater concern seems to me to be how far the localization of energy and its tracing from point to point makes any sense and significance, given our current knowledge of it. Such considerations have not yet been carried out for the simplest energy conversions of ordinary mechanics; therefore the question as to whether and to what extent the concept of energy permits such a method of treatment remains unanswered."

### 3.3 Entropy

Regarding the entropy, the development was similar but even more complicated. Also for the quantity introduced by Clausius in 1850, only a weak balance statement was formulated at first: the entropy of a closed system or the total entropy of several systems remains constant in time or increases. In an article published in 1862 Clausius writes:

> "The algebraic sum of all transformations occurring at any change of state can only be positive or zero as limit case."

For what Clausius calls a "transformation", he later introduces the name "entropy". Obviously, his proposition does not suggest the possibility to formulate a local balance law.

## 4 The breakthrough – strong balance laws

Thus, in the second half of the 19th century, three of the known balance laws had a flaw: they were only weak balance laws. However, the happy end was near: Within only 30 years all of them could be upgraded into local laws. There was a real breakthrough around the turn of the century.

Again I consider this phase of development for our three quantities one by one. However, this time I change the order, since the energy was the first of the three for which a local balance was formulated.

### 4.1 Energy

The expectation that energy balances could be described locally was already pronounced by Planck. It was finally achieved, around the time when Hertz had still formulated his concerns: first in 1884 by Poynting and in 1898 in a comprehensive article by Gustav Mie, whom we want to quote here:

> "Theorem of the continuity of energy.
>
> All energy transfers are the consequences of real energy flows."

### 4.2 Momentum

Here, things are more complicated. The problem was not the transmission of momentum through matter, but through the fields.

Actually, it could have been solved with the emergence of the first field theory of physics, the theory of electromagnetism by Faraday and Maxwell.

In 1864 Maxwell (Maxwell 1954) published his *Treatise on Electricity and Magnetism*, in which he introduced what is now known as the Maxwell stress tensor. Within an electromagnetic field, there is mechanical stress, and Maxwell showed how to calculate this stress from the field strengths:

> "105.] If the action of $E_2$ on $E_1$ is effected, not by direct action at a distance, but by means of a distribution of stress in a medium extending continuously from $E_2$ to $E_1$, it is manifest that if we know the stress at every point of any closed surface $s$ which completely separates $E_2$ from $E_1$, we shall be able to determine completely the mechanical action of $E_2$ on $E_1$. For if the force on $E_1$ is not completely accounted for by the stress through $s$, there must be direct action between something outside of $s$ and something inside of $s$.
>
> Hence if it is possible to account for the action of $E_2$ on $E_1$ by means of a distribution of stress in the intervening medium, it must be possible to express this action in the form of a surface-integral extended over any surface $s$ which completely separates $E_2$ from $E_1$."

That could have been the moment to throw Newton's provisional wording overboard: From now on, the force could have been called momentum current strength, and the mechanical stress would be the momentum current density. But apparently, nobody got aware of this possibility at that time. The old language, which was still tailored to actions at a distance, was firmly established, and those who used it had become accustomed to the flaw of the underlying idea of actions at a distance. So, we can say that the mathematics for the local description of the momentum balance was there. What was missing, was the wording, and the corresponding mental representation.

The insight came only a few years after the corresponding step was done with the energy. In 1908 Max Planck, who 21 years earlier had already demanded the local formulation of energy conservation, also formulated the conservation of momentum by means of the concept of momentum currents. In a short essay in the *Physikalische Zeitschrift* he writes:

> „Just as the constancy of energy involves the concept of energy flow, so also does the constancy of the quantity of motion necessarily involve the concept of the 'flow of the quantity of motion', or to put it more briefly: the 'momentum flow'."

### 4.3 Entropy

Finally, entropy was still missing. In 1911, the same year in which Callendar showed that the entropy introduced by Clausius can be identified with Carnot's Caloric, the entropy current also appeared on stage; at first somewhat hidden in an extensive, somewhat difficult to read publication by G. Jaumann (1911), and shortly afterwards in a paper by his scholar Lohr (1916). Lohr expresses the entropy balance exactly in the form of equation (3).

## 5 How local balancing was received

Within less than 30 years, energy, momentum, and entropy had turned into quantities for which a local balance could be established.

The genesis of the idea of local balancing of the three considered quantities was long and arduous, but the result was simple. All three quantities are simpler in structure than one had first suspected.

But how do we deal with them today? Do we benefit from the advantages of local balancing? Not at all. It is as if the state of that part of physics of the year 1890 had been frozen.

Again we consider the three quantities one after the other.

### 5.1 Momentum

The benefits of introducing momentum as a quantity for which a local balance can be established are:

1. It brings mechanics into a form that corresponds to our current requirements on a theory: The language of momentum current mechanics no longer contains formulations that suggest actions at a distance.

2. It brings mechanics into a form that has the same structure as other areas of physics, namely electricity and thermodynamics (Herrmann and Schmid 1985a).

The transition from the force model to the momentum current model also causes new challenges. Because in momentum current mechanics the wording and the thinking associated with it are tailored to a local view questions are suggested which do not arise in the force model. If we say that momentum goes or flows from A to B, an obvious question is how it gets from A to B, and that means: How is the distribution of the momentum current density. Those who engage in the momentum current representation must therefore reckon with these questions (Herrmann and Schmid 1984, Herrmann and Schmid 1985b, Heiduck et al. 1987, Grabois and Herrmann 2000). In Newton's force formulation they do not show up. They are "swept under the carpet".

However, the corresponding paradigm shift was not possible. The mental images associated with the Newtonian language have become firmly established. More than 100 years after Planck's proposition to interpret forces as momentum currents only few textbooks, almost all of them for advanced learners, use this approach, see for example Feynman (2006) or Landau-Lifshiz (1987).

Only few attempts have been made to introduce it into that teaching where the benefit would be greatest: in texts for beginners (diSessa 1980, Herrmann 2000).

### 5.2 Energy

As far as energy is concerned, the situation is slightly better: Everyone knows that one can establish energy balances, that energy is transported, that one can speak of an energy density, an energy flow and the distribution of the flow in space, i.e. an energy flow density.

And yet: Old ways of speaking and the images associated with them have survived. Instead of: "In a bicycle chain the energy flows together with momentum from the front sprocket to the rear", one uses to say: "the front sprocket wheel does work at the rear wheel". Instead of saying, the electric motor delivers an energy current of 200 W, we say its power is 200 W. Instead of saying that the radiation that the sun is emitting carries an energy current of $3.8 \cdot 10^{26}$ W, we say its luminosity is $3.8 \cdot 10^{26}$ W.

This way of speaking goes back to the time when a local balancing of the energy was not yet possible or even to the time when the physical quantity energy did not yet exist.

### 5.3 Entropy

Here, the state of teaching is farthest from a modern local causes view. The term entropy current is hardly used in physics textbooks. However, there are exceptions, see for instance Job and Rüffler (2011 chapter 3) or Fuchs (chapter 4).

## 6 What is to be done?

The answer to this question is simple: Introduce and treat energy, momentum and entropy as quantities for which a local balance can be established and use the appropriate language.

When dealing with energy in this way, one will typically ask questions like these:

How much energy is contained in the system under consideration? Where is the energy? Where is it coming from, where is it going? Which way does it go? These are questions for the distribution of the energy density and the energy current density.

The corresponding questions should be asked when dealing with momentum and entropy. Because of the validity of the local balance equations, they always have clear answers.